\input{aipcheck}

\documentclass[
    ,final            
  ]
  {aipproc}

\layoutstyle{6x9}

\begin{document}

\title{The Influence of AGN Feedback on Galaxy Cluster Observables}


\author{H.-Y. Yang}{
  address={Department of Astronomy, University of Illinois at Urbana-Champaign, Urbana, IL 61801}
}

\author{P.M. Ricker}{
  address={Department of Astronomy, University of Illinois at Urbana-Champaign, Urbana, IL 61801},
  altaddress={National Center for Supercomputing Applications, University of Illinois at Urbana-Champaign, Urbana, IL 61801}
}

\author{P.M. Sutter}{
  address={Department of Physics, University of Illinois at Urbana-Champaign, Urbana, IL 61801-3080}
}

\begin{abstract}
 Galaxy clusters are valuable cosmological probes. However, cluster mass estimates rely on observable quantities that are affected by complicated baryonic physics in the intracluster medium (ICM), including feedback from active galactic nuclei (AGN). Cosmological simulations have started to include AGN feedback using subgrid models. In order to make robust predictions, the systematics of different implementations and parametrizations need to be understood. We have developed an AGN subgrid model in FLASH that supports a few different black hole accretion models and feedback models. We use this model to study the effect of AGN on X-ray cluster observables and its dependence on model variations.
\end{abstract}

\maketitle


\section{Introduction}

Clusters of galaxies are useful probes of cosmological parameters, provided that their masses can be determined accurately from observables such as X-ray luminosity or X-ray temperature. However, it is still a challenge for current theoretical models to reproduce all the observed ICM properties. Feedback from active galactic nuclei (AGN) has been commonly invoked as the necessary missing piece. Since AGN feedback involves a wide range of scales, from the accretion disk on AU scales to clusters on Mpc scales, only recently has it been included in cosmological simulations using subgrid modeling.

Because there are a variety of implementations and parametrizations of AGN subgrid models in the literature, the aim of this study is to understand the current theoretical uncertainties in predicting the ICM properties. More specifically, we compare two black hole (BH) accretion models and two feedback models that are commonly used: bubble feedback \cite{Sijacki} and jet feedback \cite{Cattaneo}. We will describe the analytical and numerical approaches in the following section. In Section 3, we will discuss the effect of AGN feedback on cluster observables with these model variations.

\section{Methodology}

We performed hydrodynamic simulations with radiative cooling and AGN feedback with an isolated cluster sitting in a 2048\ kpc box using FLASH\ 2.5 \cite{Flash}. The region surrounding a central $10^7\ h^{-1}\mbox{M}_\odot$ black hole is refined to the maximum resolution of 4\ kpc. Following [4], the cluster gas is initialized assuming a modified NFW \cite{NFW} profile and is in hydrostatic equilibrium in an NFW gravitational potential. The cluster has a mass of $10^{14}\ h^{-1} \mbox{M}_\odot$, concentration of 6.5, cool-core radius of 24.5\ kpc, and gas fraction of 0.12. Radiative cooling is computed using \cite{SutherlandDopita} assuming zero metallicity. The Hubble constant $h=0.7$ is used.

We relate the BH accretion rate to the Bondi-Hoyle-Lyttleton \cite{Bondi} accretion rate:
\begin{equation}
\dot{M}_{\mathrm{BH}} \propto \dot{M}_{\mathrm{Bondi}} = 4\pi G^2 M^2_{\mathrm{BH}} \rho / c_{\mathrm{s}}^3,
\end{equation}
where $M_{\mathrm{BH}}$ is the BH mass, and $\rho$ and $c_{\mathrm{s}}$ are the gas density and sound speed, respectively. The proportionality reflects the fact that cosmological simulations usually do not have sufficient resolution to resolve the Bondi radius, and hence using only the Bondi accretion rate would underestimate the actual BH accretion rate. There are a number of ways to link these two quantities. In this study we will present results using the constant-$\alpha$ model, $\dot{M}_{\mathrm{BH}} = \alpha \dot{M}_{\mathrm{Bondi}}$ \cite{Sijacki}, and the density-dependent $\beta$ model \cite{Booth}:
\begin{equation}
\dot{M}_{\mathrm{BH}} = (n_H/n_H^*)^\beta \dot{M}_{\mathrm{Bondi}},
\end{equation}
where $n_H^* = 0.1\ \mathrm{cm}^{-3}$. We impose an upper limit on the accretion rate corresponding to the Eddington rate, $\dot{M}_{\mathrm{Edd}} = (4\pi G M_{\mathrm{BH}} m_{\mathrm{p}}) / (\epsilon_{\mathrm{f}} \sigma_{\mathrm{T}} c)$, where $m_{\mathrm{p}}$ is the mass of the proton, $\sigma_{\mathrm{T}}$ is the Thompson cross-section and $\epsilon_{\mathrm{f}} $ is the radiative efficiency.  

The feedback from the AGN to the surrounding gas is linked to the BH accretion rate. For bubble feedback, as in \cite{Sijacki}, the bubbles are only formed when the BH mass increases by a fraction $\delta_{\mathrm{BH}}$ since the last bubble formation. When a bubble is formed, only thermal energy is injected:
\begin{equation}
E = \epsilon_{\mathrm{m}} \epsilon_{\mathrm{f}} c^2 \delta M_{\mathrm{BH}},
\end{equation}
where $\epsilon_{\mathrm{m}}$ is the efficiency of mechanical heating, and $\delta M_{\mathrm{BH}} \equiv \delta_{\mathrm{BH}} M_{\mathrm{BH}}$ is the increase in BH mass since the last bubble was formed. The injected energy is evenly distributed within a sphere of radius, 
\begin{equation}
R = R_0 \left( \frac{E/E_0}{\rho/\rho_0} \right)^{1/5}, 
\end{equation}
where the scaling parameter values $R_0 = 30\ h^{-1} \mbox{kpc}$, $E_0 = 5\times 10^{60}$ erg, and $\rho_0 = 10^6\ h^2 \mbox{M}_\odot \mbox{kpc}^{-3}$ are motivated by observed bubble sizes. The bubbles are centered on the black hole. The default bubble parameters are $\alpha=350$, $\beta=2$, $\epsilon_{\mathrm{f}} = 0.1$, $\epsilon_{\mathrm{m}} = 1$, and $\delta_{\mathrm{BH}} = 0.001$.

For jet feedback, we adopt the model in \cite{Cattaneo}, where the injection rates of the mass, momentum, and energy are given by
\begin{eqnarray}
\dot{M} &=& \eta \dot{M}_{\mathrm{BH}} |\Psi|, \\
\dot{\bf{P}} &=& \sqrt{2\epsilon_{\mathrm{f}}} \dot{M}_{\mathrm{BH}} c \Psi, \\
\dot{E} &=& \epsilon_{\mathrm{f}} \dot{M}_{\mathrm{BH}} c^2 |\Psi|,
\end{eqnarray}
where $\eta$ is the mass loading factor, and the function $\Psi$ determines the spatial extent of the jet:
\begin{equation}
\Psi ({\bf x}) = \frac{1}{2\pi r^2_{\mathrm{ej}}} \exp \left( - \frac{x^2+y^2}{2r^2_{\mathrm{ej}}} \right) \frac{z}{h^2_{\mathrm{ej}}}.
\end{equation}
The jet is aligned with the $z$-axis and the feedback is applied to regions with $|z| \leq h_{\mathrm{eq}}$ and $r \leq 2.6 r_{\mathrm{ej}}$. The default values of the jet parameters are $\eta = 100$, $\epsilon_{\mathrm{f}} = 0.1$, $r_{\mathrm{ej}} = $3.2\ kpc, and $h_{\mathrm{ej}} =$ 2.5\ kpc. 

\section{Results}  

Figure 1 shows the growth of the black hole and its accretion history for different accretion and feedback models. We can see that the overall behavior is similar: the black hole accretion rate (BHAR) is regulated after $t \sim 0.5$ Gyr, which leads to a converged BH final mass that is in general consistent with observed BH masses in $10^{14} h^{-1} \mbox{M}_\odot$ clusters. However, the paths of BH growth are different, namely, jet feedback results in a more gradual increase in BH mass, while for the bubble feedback the black hole grows more rapidly initially and reaches a quasi-static state earlier than in the jet model. This difference is mainly driven by the behavior of the accretion rate at the early stage ($t < 1$ Gyr). Since the bubbles are only allowed to form when the BH mass exceeds the threshold fraction $\delta_{\mathrm{BH}}$, the black hole is able to accrete more before feedback begins to dominate. In contrast, the continuous jet feedback slows down the accretion from the beginning and thus yields a more mild BH accretion and mass growth. Similarly, the black hole in the $\beta$ model grows a little more slowly than in the $\alpha$ model due to the difference in the initial estimate of the accretion rate. 

\begin{figure}[thbp]
\includegraphics[height=0.26\textheight]{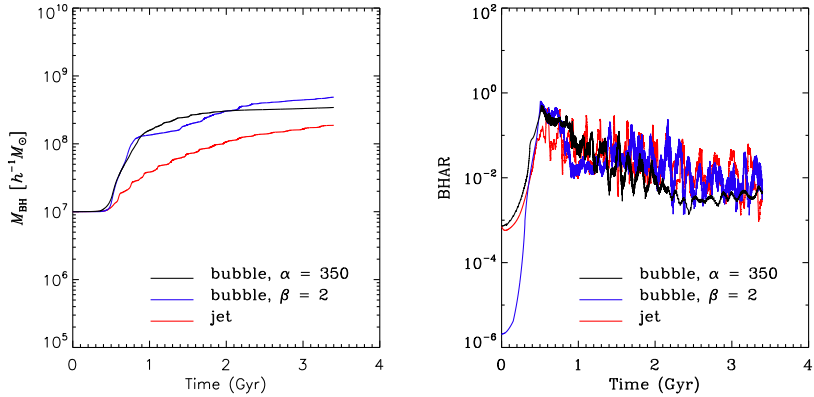}
\caption{Evolution of BH mass (left) and BH accretion rate in units of the Eddington rate (right) for different accretion and feedback models.}
\label{accre}
\end{figure}

\begin{figure}[thbp]
\includegraphics[height=0.26\textheight]{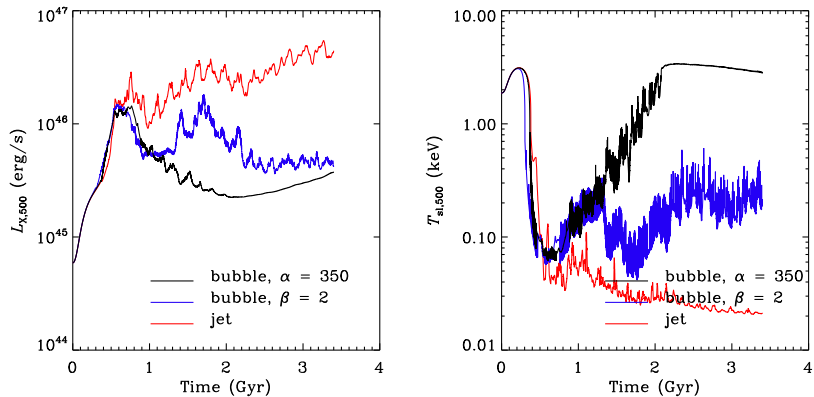}
\caption{Evolution of X-ray luminosity (left) and spectroscopic-like temperature (right) for different accretion and feedback models.}
\label{accre}
\end{figure}

In Fig.\ 2 we show the evolution of cluster X-ray observables. We can see that different accretion and feedback models actually can have quite different results. We find that in terms of heating the ICM, the $\alpha$ model with bubble feedback is most effective and the jet feedback is least effective. One possible reason is that since the amount of injected energy is proportional to the BH mass, less feedback energy is available for the jet model, in which the black hole grows more slowly. Moreover, only a portion of the feedback energy is turned into thermal energy in the jet model, while in the current setup for bubble feedback, all the energy is used to heat the cluster. Using a more realistic $\epsilon_{\mathrm{m}}$ may reduce the discrepancy. 

In conclusion, using a set of isolated cluster simulations we find a non-negligible difference in predicting the growth of black hole and influence on cluster observables using different subgrid AGN models. Model uncertainties should be carefully examined in future cosmological simulations with AGN feedback.



\begin{theacknowledgments}
The authors acknowledge support under a Presidential Early Career Award from the U.S. Department of Energy, Lawrence Livermore National Laboratory (contract B532720). Additional support was provided by NASA Headquarters under the NASA Earth and Space Science Fellowship Program (NNX08AZ02H), a DOE Computational Science Graduate Fellowship (DE-FG02-97ER25308), and the National Center for Supercomputing Applications. The software used in this work was in part developed by the DOE-supported ASC / Alliance Center for Astrophysical Thermonuclear Flashes at the University of Chicago.
\end{theacknowledgments}




\bibliographystyle{aipprocl} 




\end{document}